\documentclass[aps,prd,%
onecolumn,%
tightenlines,%
preprint,
eqsecnum,%
showpacs,%
floats,%
nofootinbib,%
amsfonts,amssymb,amsmath%
]{revtex4-1}

\newcommand{\ket}[1]{\mbox{$| {#1} \rangle$}}
\newcommand{\bracket}[2]{\mbox{$\langle {#1} \!\mid\! {#2} \rangle$}} 
\newcommand{\ketbra}[2]{\mbox{$| {#1} \rangle\langle {#2} |$}}

\newcommand{\Projsupb}[2]{\mbox{$P^{{#1}}_{\mbox{\scriptsize{${#2}$}}}$}}

\def\hilbert{{\mathcal H}}

\begin{document}


\title{Consistent Histories in Quantum Cosmology}

\author{David Craig}
\email[]{E-mail: craigda@lemoyne.edu}
\affiliation{%
Department of Physics, Le Moyne College\\
Syracuse, New York, 13214, USA}

\author{Parampreet Singh}
\email[]{E-mail: psingh@perimeterinstitute.ca}
\affiliation{%
Perimeter Institute for Theoretical Physics\\
Waterloo, Ontario, N2L 2Y5, Canada}

\date{\today}

\begin{abstract}
We illustrate the crucial role played by decoherence (consistency of
quantum histories) in extracting consistent quantum probabilities for
alternative histories in quantum cosmology.  Specifically, within a
Wheeler-DeWitt quantization of a flat Friedmann-Robertson-Walker
cosmological model sourced with a free massless scalar field, we
calculate the probability that the univese is singular in the sense
that it assumes zero volume.  Classical solutions of this model are a
disjoint set of expanding and contracting singular branches.  A naive
assessment of the behavior of quantum states which are superpositions
of expanding and contracting universes may suggest that a ``quantum
bounce'' is possible \emph{i.e.\ }that the wave function of the
universe may remain peaked on a non-singular classical solution throughout
its history.  However, a more careful consistent histories analysis
shows that for arbitrary states in the physical Hilbert space the
probability of this Wheeler-DeWitt quantum universe encountering the
big bang/crunch singularity is equal to unity.  A quantum
Wheeler-DeWitt universe is inevitably singular, and a ``quantum
bounce'' is thus not possible in these models.
\end{abstract}

\pacs{98.80.Qc,03.65.Yz,04.60.Ds,04.60.Kz}  
%

\maketitle

\section{Introduction}
\label{sec:intro}

Whatever form a theory of quantum gravity may ultimately take -- 
string theory, loop quantum gravity, causal dynamical triangulations, causal
sets, or something else%
\footnote{See \cite{kiefer07} for an overview.
} %
-- extracting coherent physical predictions will require more than just the
basic technical apparatus of the theory.  Because one of the explicit aims of
such theories is to provide a complete quantum description of the universe
\emph{as a whole}, a framework within which consistent predictions can be made
in the absence of external observers or an external notion of ``measurement''
will also be required in an essential way.

In this essay we aim to outline an explicit realization of one such
framework, the generalized ``consistent histories'' quantum theory of
J.B.\ Hartle \cite{LesH}, in the context of the Wheeler-DeWitt
quantization of a flat homogeneous and isotropic
Friedman-Robertson-Walker (FRW) model universe populated with a free,
massless, minimally coupled scalar field.  (Previous discussions of
decoherent histories formulations of quantum cosmological models
include
\cite{hartlemarolf97,hallithor02,CH04,as05,halliwall06,halliwell09}.)
We will describe when and how, within the framework of decoherent
histories, consistent quantum predictions may be extracted from the
theory -- namely, when the histories describing the possible outcomes
decohere -- emphasizing the fundamental quantum fact that it is
\emph{not} normally the case that a probability may sensibly be
assigned to every history of a system.
We illustrate with an example of a prediction that depends crucially
on decoherence of the appropriate histories: whether or not
Wheeler-DeWitt quantized FRW universes are singular.  The classical
solutions of this model are all singular and fall into disjoint sets
of ever-expanding or ever-contracting universes.  We shall show that
decoherence is critical to a coherent \emph{quantum} analysis of this
question, concluding unambiguously within this framework that all
quantum Wheeler-DeWitt FRW universes are singular as well.  This
stands in contrast to a naive analysis of the behavior of the quantum
wave function alone, not taking care of the proper description and
decoherence of the corresponding histories.  This (erroneous) analysis
of states which are superpositions of wavefunctions peaked on
classically disjoint expanding and contracting solutions seems to
suggest that quantum avoidance of the singularity \emph{is} possible.
However, a careful analysis of the quantum histories shows that the
probability of avoidance of the singularity is zero and the
probability that a Wheeler-DeWitt quantum universe encounters a
singularity is unity.

The organization of this paper is as follows.  Section \ref{sec:gqt}
describes the overall framework of generalized quantum theory.  In
section \ref{sec:model} we give details of the cosmological model and
its (canonical) quantization, including a discussion of the model's
observables, and in section \ref{sec:sing} we discuss the analysis of
the quantum singularity.  (Complete technical details of the model and
its decoherent histories formulation will be given elsewhere
\cite{CS10a}.)

\section{Generalized ``Consistent Histories'' Quantum Theory}
\label{sec:gqt}

Quite generally, the job of any quantum theory is to do two things.  For any 
given physical system, a quantum theory must, for all observable quantities, 
tell both \emph{What are the possible values of those observables?} and 
\emph{How likely is each of these values?}  
In Hilbert space formulations of quantum theory, where observables are
represented by self-adjoint operators $\{\hat{A}\}$ on a Hilbert space
$\hilbert$, the answer to the first question is found in the eigenvalues
$\{a\}$ of those operators, and to the second, in amplitudes constructed from
the corresponding eigenstates $\{\ket{a}\}$ and the state $\ket{\Psi(t)}$ of
the system at the moment in question:
\begin{equation}
p_a= |\bracket{a}{\Psi(t)}|^2.
\label{eq:proba}
\end{equation}
In the textbook ``Copenhagen'' view, this framework is usually supplemented
with the rule that once a particular outcome $\{a'\}$ has been ``observed'',
the state of the system is replaced by that of the observed outcome,
$\ket{\Psi(t)}\rightarrow \{\ket{a'}\}$.   This is commonly referred to as 
the ``collapse of the wave function''.

But when does it make sense to assign such probabilities, and thence reassign
the state of the system in this discontinuous, uncontrollable way?  Certainly
such amplitudes do not \emph{always} make sense.   Most commonly, this is the 
case when considering the likelihood of \emph{sequences} of outcomes, as is 
well known, for example, in the two-slit experiment \cite{GZ05}.  Quite 
generally, in experiments of this kind, probabilities may be consistently 
assigned \emph{only} when there are physical mechanisms in place (such as a 
measuring apparatus or a suitable physical environment) which \emph{destroy 
the interference} between the possible alternative histories.  

Taking a rather conservative approach, this idea has been formalized into a
scheme known as the ``decoherent'' or ``consistent'' histories formulation of
quantum theory, in which probabilities may be consistently assigned
\emph{whenever} interference between histories vanishes, irrespective of the
presence of measuring devices, external observers, or other such notions; see
\cite{Omnes94,LesH,Griffiths08} for broad overviews.  In fact, the principal
virtue of this perspective is that it employs criteria \emph{entirely
internal} to the system 
to determine whether or not the assignment of probabilities makes coherent sense.
\footnote{By which it is meant that the appropriate probability sum rules are
satisfied.
} %

The most general formulation of the consistent histories program, suitable for
application to quantum cosmology, is due to J.B.\ Hartle \cite{LesH}.  In 
this form it is often referred to as ``generalized quantum theory''.   We 
summarize very briefly its chief elements for theories with a 
homogeneous time here.

Particular sequences of physical outcomes (``histories'') are characterized by
``class operators'', products of Heinseberg projections onto ranges of 
eigenvalues of observables of interest:
\begin{subequations}
\begin{eqnarray}
C_h &=& \Projsupb{\alpha_1}{\Delta a_{k_1}}(t_1)
        \Projsupb{\alpha_2}{\Delta a_{k_2}}(t_2) \cdots
        \Projsupb{\alpha_n}{\Delta a_{k_n}}(t_n) 
\label{eq:classopdefqm-a}  \\
    &=& U(t_0-t_1)\Projsupb{\alpha_1}{\Delta a_{k_1}}
        U(t_1-t_2)\Projsupb{\alpha_2}{\Delta a_{k_2}} \cdots
        U(t_{n-1}-t_n)\Projsupb{\alpha_n}{\Delta a_{k_n}}U(t_n-t_0), 
\label{eq:classopdefqm-c}
\end{eqnarray}
\label{eq:classopdefqm}%
\end{subequations}
where $U(t)=\exp(-i\hat{H}t/\hbar)$ is the system's propagator
and $t_1\leq t_2\leq \cdots \leq t_n$.  Given the
initial state of the system $\ket{\psi}$, the ``branch wave function''
corresponding to the state of a system following the history $h$ is
constructed from the class operator for $h$:
\begin{equation}
\ket{\psi_h} = C_h^{\dagger}\ket{\psi}.  
\label{eq:bwfdef}
\end{equation}
One may see that $\ket{\psi_h}$ is exactly the state one would assign to this 
system in ``Copenhagen'' quantum theory were the observed outcomes to be the 
ranges $\Delta a_{k_i}$; the projections correspond to the ``collapses''.   
In a histories-oriented formulation, however, $\ket{\psi_h}$ is not the state 
of the system, $\ket{\psi}$ is.   The apparent discontinuous evolution of the 
state in Copenhagen quantum theory is the result of insisting on assigning a 
meaning to $\ket{\psi(t)}$ \emph{separately along every possible ``branch''} 
of the system.

The probability of the history $h$ is given by the L\"{u}ders-von Neumann 
formula:
\begin{equation}
p(h) = \bracket{\psi_{h}}{\psi_{h}}.
\label{eq:probh}
\end{equation}
Such a probability only makes sense, however -- meaning $\sum_{h} p(h)=1$ -- 
when the interference between the possible histories vanishes, 
$\bracket{\psi_{h'}}{\psi_{h}} = 0$ for $h'\neq h$.

One therefore defines the ``decoherence functional''
\begin{equation}
d(h,h') = \bracket{\psi_{h'}}{\psi_{h}}.
\label{eq:dfdef}
\end{equation}
When the decoherence functional is diagonal on a given set of histories,%
\footnote{Complete in the sense that it consists in a set of mutually
exclusive, collectively exhaustive histories.
} %
then that set is said to ``decohere'' or ``be consistent''.  In such sets
\begin{equation}
d(h,h') = p(h)\, \delta_{h'h}.
\label{eq:ffqm}
\end{equation}
The decoherence functional of a system thus provides an internally defined,
objective measure of interference between the possible histories of a system,
independent of notions of external observers or measurements -- though it does
reproduce the predictions of ordinary quantum theory in appropriate
``measurement situations''.  It is the decoherence functional that determines
when the quantum theory makes a prediction concerning a possible history, and,
when a probability may be sensibly assigned, what that probability is.

\section{Model cosmology \& its quantization}
\label{sec:model}

To illustrate the essential importance of the consistency of alternative
histories to prediction in quantum gravity, we consider the predictions of a 
simple quantum cosmological model concerning the volume of the universe.  We 
find that decoherence is \emph{critical} to a coherent prediction, and 
indeed, if one ignores the role it plays, one is led to infer a conclusion 
which is in fact the precise \emph{opposite} of the correct prediction. 
The point is not so much the applicability of the content of the prediction 
itself for our universe.   Different models might -- and indeed, do -- give
a different result \cite{CS10b}.  Rather, it is to emphasize the essential 
role played by quantum considerations in \emph{making} such predictions.

In this paper, we concentrate on the implications of these quantum
considerations, referring to \cite{CS10a} for technical details.

The model we consider is a flat, homoegeneous and isotropic 
Friedmann-Robertson-Walker (FRW) universe with line element
\begin{equation}
ds^2 = -dt^2+a^2(t)\,d\mathring{q}^2,
\label{eq:FRWmetric}
\end{equation}
where $d\mathring{q}^2$ is a fixed, flat fiducial metric on the
spatial slices.  Since the flat FRW universe is spatially infinite we
choose a fiducial cell with unit spatial volume with respect to
$d\mathring{q}^2$ and restrict all integrations to it.  Choosing for
matter content a free, massless, minimally coupled scalar field, the
action for this model cosmology is
\begin{equation}
S = \frac{3}{8\pi G}\int\!dt\,(-a\dot{a}^2) + \frac{1}{2} \int\!dt\, a^3\dot{\phi}^2.
\label{eq:FRWaction}
\end{equation}
Canonical phase space variables for the model are
$\{a,p_a,\phi,p_{\phi}\}$, the scale factor and scalar field and their
conjugate momenta.  A canonical transformation to the volume $V=a^3$
and its conjugate $\beta=-\frac{4\pi G}{3}\frac{p_a}{a^2}$ (so
$\{\beta,V\}=4\pi G$) \cite{acs:slqc} puts the Hamiltonian constraint
in the form
\begin{equation}
-\beta^2V^2 +\frac{4\pi G}{3}p_{\phi}^2 \approx 0,
\label{eq:Hconstraint}
\end{equation}
from which one sees that $\beta^2$ measures the energy density $p_{\phi}^2/2V^2$ in the classical theory. 
Classically, $p_{\phi}$ is a constant of the motion, %
and
\begin{equation}
\phi = \pm\frac{1}{\sqrt{12\pi G}}\ln\frac{V}{V_c}+\phi_c
\label{eq:classsolns}
\end{equation}
are expanding and contracting branches of the classical solution.  ($V_c$ and
$\phi_c$ are integration constants.)

It is shown in \cite{aps:improved,acs:slqc} that this model may be
canonically quantized by setting $[\hat{\beta},\hat{V}]= i 4\pi G \hbar$
and $\hat{p}_{\phi}=-i\hbar\,\partial/\partial\phi$.  Defining a
rescaled volume variable
\begin{equation}
\nu=\frac{V}{(\frac{4\pi}{3})^{3/2}\, l_p^3/K},
\label{eq:nu}
\end{equation}
where $K=2\sqrt{2}/3\sqrt{3\sqrt{3}}$, %
the constraint
becomes the Wheeler-DeWitt equation
\begin{equation}
\partial_{\phi}^2\,\chi(\nu,\phi) = -\Theta(\nu)\chi(\nu,\phi),
\label{eq:WdW}
\end{equation}
where $\Theta(\nu)=-2\pi G\,\nu\partial_{\nu}(\nu\partial_{\nu})$ is
positive definite and self adjoint on $L^2(R,\nu^{-1}d \nu)$.  General
solutions of the quantum constraint may be decomposed in terms of
positive and negative frequency parts, each of which satisfy the first
order Schr\"{o}dinger-like quantum constraint equation:
\begin{equation}
\mp i \, \partial_{\phi}\chi_{\pm}(\nu,\phi) = \sqrt{\Theta} \, \chi_{\pm}(\nu,\phi) ~.
\label{eq:schroed}
\end{equation}

The scalar field $\phi$ thus appears mathematically in the role of a
``time'' parameter,%
\footnote{This interpretation may provide a convenient way of
conceptualizing mathematical results, but is by no means essential to
the analysis or its conclusions.
} %
and the dynamics may be expressed in terms of the propagator
$U(\phi)=\exp(i\sqrt{\Theta}\phi)$.   

The multiplicative volume operator $\hat{\nu}$ has eigenvalues
$0\leq\nu<\infty$ with eigenfunctions satisfying
\begin{equation}
\bracket{\nu}{\nu'}=\delta(\ln\nu-\ln\nu')
\label{eq:normvol}
\end{equation}
and projections on the range $d\nu$ given by
\begin{equation}
dP_{\nu}=\frac{d\nu}{\nu}\ketbra{\nu}{\nu}.
\label{eq:projvolinf}
\end{equation}
The scalar momentum $\hat p_\phi$ is a constant of the motion just as
it is in the classical theory.  Physical predictions may be extracted
from observables which commute with the quantum constraint.  These are
the Dirac observables $\hat p_\phi$ and $\hat \nu|_\phi$.  Here $\hat
\nu_{|\phi^*}$, defined by $\hat \nu_{|\phi^*} \, \psi(\nu,\phi) =
\exp{i \sqrt{\Theta}(\phi - \phi^*)} \, \hat{\nu}\, \psi(\nu,\phi^*)$,
is the ``relational observable'' corresponding to $\hat\nu$ giving the
volume at $\phi=\phi^*$.

The inner product and hence the physical Hilbert space may  be
found by demanding that the action of these observables be self
adjoint; it turns out to be \cite{aps:improved}
\begin{equation}
\bracket{\chi}{\psi} = \int_{0}^{\infty}\!\frac{d\nu}{\nu}\chi^*(\nu,\phi)\psi(\nu,\phi) ~.
\label{eq:ip}
\end{equation}
The same inner product can be obtained from a more rigorous group
averaging procedure \cite{marolf:groupavg,almmt:groupavg}.  
(See Ref.\ \cite{aps} for details.)

Because the dynamics (and hence Dirac observables) preserve the
positive and negative frequency subspaces in the physical Hilbert
space, we can restrict to either subspace.  Here we choose to work
with positive frequency solutions.  Solutions to the Wheeler-DeWitt
equation may then be represented as a sum of ``expanding'' ($R$ for
``right-moving'' in a plot of $\phi$ \emph{vs.\ }$\nu$) and
``contracting'' ($L$ for ``left-moving'') orthogonal branches,
\begin{equation}
\Psi(\nu,\phi) = \Psi^R(\nu_-) + \Psi^L(\nu_+),
\label{eq:LRdecomp}
\end{equation}
where $\nu_{\pm} = \ln\nu \pm \sqrt{12\pi G}\,\phi$.  (Negative
frequency solutions can be written in a similar way).  

Defining Heisenberg projections with the dynamics $U(\phi)$, class
operators for histories pertaining to values of physical quantities at
specific values of $\phi$ may be defined similarly to Eq.\
(\ref{eq:classopdefqm}).  The decoherence functional for this quantum
cosmological model may then be defined just as in Eq.\
(\ref{eq:dfdef}) with the inner product given by Eq.\ (\ref{eq:ip})
\cite{CH04}.%
\footnote{The striking similarity of this formulation of generalized 
quantum theory for our cosmological model with that of ordinary 
non-relativistic quantum mechanics described in section \ref{sec:gqt} 
is a consequence of two factors.   First, in this quantization a 
natural ``clock'' (namely, the scalar field $\phi$) appears in the 
model through Eq.\ (\ref{eq:schroed}).  Second, we are here restricting 
consideration to histories of alternatives defined by sequences of 
values of physical quantities at definite values of that clock.   
More general categories of alternatives, and therefore class 
operators, are certainly of potential physical relevance.   We do not 
however consider them here.   
Previous discussions of decoherent histories formulations of quantum 
cosmological models may be found in
\cite{hartlemarolf97,hallithor02,CH04,as05,halliwall06,halliwell09}.
These prior formulations generally lack, however, a well-defined 
underlying Hilbert space and observables.  On the other hand, some of 
them may be better suited to models with no emergent ``clock''.
} %

One can then show within the framework of generalized quantum theory 
outlined above that the probability (for the history in which) the volume 
$\nu$ is in a range $\Delta\nu$ when $\phi=\phi^*$ is given by%
\begin{equation}
p_{\Delta\nu}(\phi^*) = \int_{\Delta\nu}\frac{d\nu}{\nu}|\Psi(\nu,\phi^*)|^2.
\label{eq:probvol}
\end{equation}

A comment concerning this result is in order.  The operator
$\hat{\nu}$ does \emph{not} commute with the quantum constraint, and is
therefore not a Dirac observable \cite{Rovelli:obs,Marolf:obs,Dittrich:obs}.
However, the corresponding relational observable $\hat\nu_{|\phi^*}$  is.  It is
satisfying that the probability for the history in which $\hat{\nu}$ assumes
values in $\Delta\nu$ at $\phi^*$ gives precisely the result one would expect
from considering expectation values of $\hat \nu_{|\phi^*}$.  One thus may see
how consideration of relational observables arises naturally within histories
formulations of quantum theory.  (Alternately, one may consider class 
operators constructed directly from projections onto ranges of values 
of the Dirac observables.)
For further details see \cite{CS10a}.%
\footnote{An alternative approach demands that the class operators 
themselves must commute with the constraint 
\cite{hartlemarolf97,hallithor02,halliwall06,halliwell09}.  
} %

\section{Quantum Singularities}
\label{sec:sing}

We turn now to the question of whether this model quantum universe is 
singular, just as the classical model is.   There are many incommensurate 
criteria one might apply to this question.   We focus on a simple one: does 
the universe ever achieve zero volume?

For a universe whose state is purely ``left-moving'' (contracting) or 
``right-moving'' (expanding) (\emph{cf.\ }Eq.\ (\ref{eq:LRdecomp})), this 
question is easy to answer in the affirmative.  From Eq.\ (\ref{eq:probvol}) 
it is straightforward to show that if $\Delta\nu^*=[0,\nu^*]$ 
and$\ket{\Psi}=\ket{\Psi^L}$ or $\ket{\Psi^R}$,
\begin{subequations}
\begin{align}
\lim_{\phi\rightarrow-\infty} p^L_{\Delta\nu^*}(\phi) &= 0
&
\lim_{\phi\rightarrow+\infty} p^L_{\Delta\nu^*}(\phi) &= 1
\label{eq:label-a}\\
\lim_{\phi\rightarrow-\infty} p^R_{\Delta\nu^*}(\phi) &= 1
&
\lim_{\phi\rightarrow+\infty} p^R_{\Delta\nu^*}(\phi) &= 0
\label{eq:label-b}
\end{align}
\label{eq:label}%
\end{subequations}
for any fixed choice of $\nu^*$, no matter how small.  This is just as one may
have expected: contracting universes will inevitably shrink to arbitrarily
small volume, and expanding universes will equally inevitably have grown
\emph{from} arbitrarily small volume.

For this prediction, the role of decoherence was essentially trivial since 
decoherence for alternatives defined only at a single moment of ``time'' 
($\phi$) is automatic.%
\footnote{This is because the corresponding class operators are simply 
projections; the branch wave functions for different histories are thereby
automatically orthogonal.
} %

Far more interesting, however, is the general case in which the state of the 
universe is a superposition
\begin{equation}
\ket{\Psi} = \ket{\Psi^L} + \ket{\Psi^R}
\label{eq:cat}
\end{equation}
of expanding and contracting universes.  Here, $p_L=\bracket{\Psi^L}{\Psi^L}$
and $p_R=\bracket{\Psi^R}{\Psi^R}$ (with $\bracket{\Psi^L}{\Psi^R} = 0$ and
$p_L + p_R =1$) measure the ``amount'' of each in the superposition.  One may 
ask, what is the likelihood that a universe in such a ``Schr\"{o}dinger's 
Cat''-like superposition of (possibly macroscopic) states could ``jump'' from 
the contracting to the expanding branch \emph{i.e.\ }remain peaked on a large 
classical solution at both ``early'' and ``late'' values of $\phi$?

Indeed, such a ``quantum bounce'' may at first seem possible.  As before, one 
may show that
\begin{equation}
p_{\Delta\nu^*}(\phi) = 
  p^L_{\Delta\nu^*}(\phi) + p^R_{\Delta\nu^*}(\phi).
\label{eq:pvolcat}
\end{equation}
Now, though,
\begin{equation}
\lim_{\phi\rightarrow -\infty} p_{\Delta\nu^*}(\phi) = p_R
\qquad \mathrm{and}\qquad
\lim_{\phi\rightarrow +\infty} p_{\Delta\nu^*}(\phi) = p_L.
\label{eq:pvolcatlim}
\end{equation}
Thus, there is in general a non-zero probability for the universe to be
non-singular in \emph{both} the ``past'' ($p_L$) and the ``future'' ($p_R$).
Is this not precisely the possibility (with probability $p_L\cdot p_R$) of a
``quantum bounce'' we were seeking?

The answer is emphatically \textbf{no}.  We are being misled because Eq.\
(\ref{eq:pvolcat}) is the answer to a specific quantum question: \emph{What is
the probability that the volume of the universe is in $\Delta\nu^*$ at a given
value of $\phi$?} But the question of whether a quantum universe bounces is
really about \emph{two} values of $\phi$: \emph{What is the probability that
the universe is \textbf{not} in $\Delta\nu^*$ at \emph{both}
$\phi\rightarrow-\infty$ \emph{and} $\phi\rightarrow +\infty$?} The class
operator for this question is
\begin{equation}
C_{\mathrm{bounce}} = 
\lim_{\substack{\phi_1\rightarrow -\infty\\ \phi_2\rightarrow +\infty }}
\Projsupb{\nu}{\overline{\Delta\nu^*_1}}(\phi_1)
\Projsupb{\nu}{\overline{\Delta\nu^*_2}}(\phi_2),
\label{eq:Cbounce}
\end{equation}
where the intervals $\overline{\Delta\nu^*}$ are the complements of the 
intervals $\Delta\nu^*$ specifying small volume.   The complementary history 
\begin{equation}
C_{\mathrm{sing}} = 1- C_{\mathrm{bounce}} 
\label{eq:Csing}
\end{equation}
is the one for which the universe is inevitably singular \emph{i.e.\ }assumes 
zero volume (enters $\Delta\nu^*$) at some point in its history.

Now, however, decoherence is \emph{not} automatic, and it is \emph{not} 
trivial.  In general, questions about the volume of the universe at two 
different values of $\phi$ make no more quantum sense than the question of 
which slit the particle passed through in two-slit interference.   The 
question can only be given a coherent answer when the histories decohere, 
such as when there is a measuring device in place.  

In this particular case, however, it is possible to show with complete
generality that \emph{in the given limit $\phi\rightarrow-\infty$ and
$\phi\rightarrow +\infty$}, the histories $\{\mathrm{bounce},\mathrm{sing}\}$ 
\emph{do} decohere.   Indeed, in this limit
%
\begin{equation}
\ket{\Psi_{\mathrm{bounce}}} = C_{\mathrm{bounce}}^{\dagger}\ket{\Psi} = 0
\qquad \mathrm{and} \qquad
\ket{\Psi_{\mathrm{sing}}} 
= C_{\mathrm{sing}}^{\dagger}\ket{\Psi} = \ket{\Psi},
\label{eq:branchvol}
\end{equation}
so that
\begin{equation}
p_{\mathrm{bounce}}= 0 \qquad \mathrm{and} \qquad p_{\mathrm{sing}}= 1.
\label{eq:probsing}
\end{equation}
We have thus shown, within an explicit framework for computing quantum 
probabilities in an objective, observer-independent way, that a quantum 
bounce is not possible in these models: the universe is inevitably singular.

\section{Discussion}
\label{sec:discuss}

What is important to underscore is the absolutely \emph{crucial} role
played by decoherence in this analysis.  A naive assessment of the
behavior of $\ket{\Psi(\phi)}$ alone (as in Eq.\ (\ref{eq:pvolcat}))
seems to imply that a quantum bounce \emph{is} possible.  This,
however, is critically misleading.  The more careful analysis that
recognizes that a quantum bounce is fundamentally a question
concerning properties of the universe at a \emph{sequence} of values
of $\phi$ shows that it is inevitably a quantum question, and must
have a quantum answer.  Such an answer is \emph{only} available if the
corresponding histories decohere.  We have shown that, in a
certain limit, they indeed do,%
\footnote{In general, these histories \emph{do not} decohere at finite values
of $\phi$, even though in the quantum theory the left- and 
right-moving sectors are orthogonal and preserved by the theory's 
Dirac observables \cite{CS10a}.
} %
and that the quantum bounce suggested by Eq.\
(\ref{eq:pvolcat}) cannot in fact occur.

Put another way, ignoring the fundamental role played by quantum decoherence 
would tend to lead one to an utterly \emph{incorrect} conclusion.  Quantum 
mechanics matters, even when applied to the universe as a whole.





\begin{acknowledgments}

D.C.\ would like to thank the Perimeter Institute, where much of this
work was completed, for its repeated hospitality.  Research at the
Perimeter Institute is supported by the Government of Canada through
Industry Canada and the Province of Ontario through the Ministry of
Research \& Innovation.

\end{acknowledgments}


\bibliography{qtrf5}

\begin{thebibliography}{10}%
\makeatletter
\providecommand \@ifxundefined [1]{%
 \ifx #1\undefined \expandafter \@firstoftwo
 \else \expandafter \@secondoftwo
\fi
}%
\providecommand \@ifnum [1]{%
 \ifnum #1\expandafter \@firstoftwo
 \else \expandafter \@secondoftwo
\fi
}%
\providecommand \enquote [1]{``#1''}%
\providecommand \bibnamefont  [1]{#1}%
\providecommand \bibfnamefont [1]{#1}%
\providecommand \citenamefont [1]{#1}%
\providecommand\href[0]{\@sanitize\@href}%
\providecommand\@href[1]{\endgroup\@@startlink{#1}\endgroup\@@href}%
\providecommand\@@href[1]{#1\@@endlink}%
\providecommand \@sanitize [0]{\begingroup\catcode`\&12\catcode`\#12\relax}%
\@ifxundefined \pdfoutput {\@firstoftwo}{%
 \@ifnum{\z@=\pdfoutput}{\@firstoftwo}{\@secondoftwo}%
}{%
 \providecommand\@@startlink[1]{\leavevmode\special{html:<a href="#1">}}%
 \providecommand\@@endlink[0]{\special{html:</a>}}%
}{%
 \providecommand\@@startlink[1]{%
  \leavevmode
  \pdfstartlink
   attr{/Border[0 0 1 ]/H/I/C[0 1 1]}%
   user{/Subtype/Link/A<</Type/Action/S/URI/URI(#1)>>}%
  \relax
 }%
 \providecommand\@@endlink[0]{\pdfendlink}%
}%
\providecommand \url  [0]{\begingroup\@sanitize \@url }%
\providecommand \@url [1]{\endgroup\@href {#1}{\urlprefix}}%
\providecommand \urlprefix [0]{URL }%
\providecommand \Eprint[0]{\href }%
\@ifxundefined \urlstyle {%
  \providecommand \doi [1]{doi:\discretionary{}{}{}#1}%
}{%
  \providecommand \doi [0]{doi:\discretionary{}{}{}\begingroup
  \urlstyle{rm}\Url }%
}%
\providecommand \doibase [0]{http://dx.doi.org/}%
\providecommand \Doi[1]{\href{\doibase#1}}%
\providecommand \bibAnnote [3]{%
  \BibitemShut{#1}%
  \begin{quotation}\noindent
    \textsc{Key:}\ #2\\\textsc{Annotation:}\ #3%
  \end{quotation}%
}%
\providecommand \bibAnnoteFile [2]{%
  \IfFileExists{#2}{\bibAnnote {#1} {#2} {\input{#2}}}{}%
}%
\providecommand \typeout [0]{\immediate \write \m@ne }%
\providecommand \selectlanguage [0]{\@gobble}%
\providecommand \bibinfo [0]{\@secondoftwo}%
\providecommand \bibfield [0]{\@secondoftwo}%
\providecommand \translation [1]{[#1]}%
\providecommand \BibitemOpen[0]{}%
\providecommand \bibitemStop [0]{}%
\providecommand \bibitemNoStop [0]{.\EOS\space}%
\providecommand \EOS [0]{\spacefactor3000\relax}%
\providecommand \BibitemShut [1]{\csname bibitem#1\endcsname}%
\bibitem{kiefer07}%
  \BibitemOpen
  \bibfield{author}{%
  \bibinfo {author} {\bibfnamefont{C.}~\bibnamefont{Kiefer}},\ }%
  \emph{\bibinfo {title} {Quantum gravity}},\ \bibinfo {edition} {2nd}\ ed.\
  (\bibinfo {publisher} {Oxford University Press, Oxford},\ \bibinfo {year}
  {2007})%
  \bibAnnoteFile{NoStop}{kiefer07}%
\bibitem{LesH}%
  \BibitemOpen
  \bibfield{author}{%
  \bibinfo {author} {\bibfnamefont{J.~B.}\ \bibnamefont{Hartle}},\ }%
  in\ \emph{\bibinfo {booktitle} {Gravitation and Quantizations, Proceedings of
  the 1992 Les Houches Summer School}},\ \bibinfo {editor} {edited by\ \bibinfo
  {editor} {\bibfnamefont{B.}~\bibnamefont{Julia}}\ and\ \bibinfo {editor}
  {\bibfnamefont{J.}~\bibnamefont{Zinn-Justin}}}\ (\bibinfo {publisher} {North
  Holland, Amsterdam},\ \bibinfo {year} {1995})\
  \Eprint{http://arxiv.org/abs/arXiv:gr-qc/9304006}{arXiv:gr-qc/9304006}%
  \bibAnnoteFile{NoStop}{LesH}%
\bibitem{hartlemarolf97}%
  \BibitemOpen
  \bibfield{author}{%
  \bibinfo {author} {\bibfnamefont{J.~B.}\ \bibnamefont{Hartle}}\ and\ \bibinfo
  {author} {\bibfnamefont{D.}~\bibnamefont{Marolf}},\ }%
  \bibfield{journal}{%
  \bibinfo {journal} {Phys.\ Rev.}\ }%
  \textbf{\bibinfo {volume} {D56}},\ \bibinfo {pages} {6247} (\bibinfo {year}
  {1997}),\ \Eprint{http://arxiv.org/abs/9703021}{arXiv:9703021 [gr-qc]}%
  \bibAnnoteFile{NoStop}{hartlemarolf97}%
\bibitem{hallithor02}%
  \BibitemOpen
  \bibfield{author}{%
  \bibinfo {author} {\bibfnamefont{J.~J.}\ \bibnamefont{Halliwell}}\ and\
  \bibinfo {author} {\bibfnamefont{J.}~\bibnamefont{Thorwart}},\ }%
  \bibfield{journal}{%
  \bibinfo {journal} {Phys.\ Rev.}\ }%
  \textbf{\bibinfo {volume} {D65}},\ \bibinfo {pages} {104009} (\bibinfo {year}
  {2002}),\ \Eprint{http://arxiv.org/abs/gr-qc/0201070}{gr-qc/0201070}%
  \bibAnnoteFile{NoStop}{hallithor02}%
\bibitem{CH04}%
  \BibitemOpen
  \bibfield{author}{%
  \bibinfo {author} {\bibfnamefont{D.~A.}\ \bibnamefont{Craig}}\ and\ \bibinfo
  {author} {\bibfnamefont{J.~B.}\ \bibnamefont{Hartle}},\ }%
  \bibfield{journal}{%
  \bibinfo {journal} {Phys. Rev.}\ }%
  \textbf{\bibinfo {volume} {D}},\ \bibinfo {pages} {123525} (\bibinfo {month}
  {June}\ \bibinfo {year} {2004}),\
  \Eprint{http://arxiv.org/abs/gr-qc/0309117v3}{arXiv:gr-qc/0309117v3}%
  \bibAnnoteFile{NoStop}{CH04}%
\bibitem{as05}%
  \BibitemOpen
  \bibfield{author}{%
  \bibinfo {author} {\bibfnamefont{C.}~\bibnamefont{Anastopoulos}}\ and\
  \bibinfo {author} {\bibfnamefont{K.}~\bibnamefont{Savvidou}},\ }%
  \bibfield{journal}{%
  \bibinfo {journal} {Class.\ Quant. Grav.}\ }%
  \textbf{\bibinfo {volume} {22}},\ \bibinfo {pages} {1841} (\bibinfo {year}
  {2005}),\ \Eprint{http://arxiv.org/abs/0410131}{arXiv:0410131 [gr-qc]}%
  \bibAnnoteFile{NoStop}{as05}%
\bibitem{halliwall06}%
  \BibitemOpen
  \bibfield{author}{%
  \bibinfo {author} {\bibfnamefont{J.~J.}\ \bibnamefont{Halliwell}}\ and\
  \bibinfo {author} {\bibfnamefont{P.}~\bibnamefont{Wallden}},\ }%
  \bibfield{journal}{%
  \bibinfo {journal} {Phys.\ Rev.}\ }%
  \textbf{\bibinfo {volume} {D73}},\ \bibinfo {pages} {024011} (\bibinfo {year}
  {2006})%
  \bibAnnoteFile{NoStop}{halliwall06}%
\bibitem{halliwell09}%
  \BibitemOpen
  \bibfield{author}{%
  \bibinfo {author} {\bibfnamefont{J.~J.}\ \bibnamefont{Halliwell}}}%
   (\bibinfo {year} {2009}),\
  \Eprint{http://arxiv.org/abs/0909.2597}{arXiv:0909.2597 [gr-qc]}%
  \bibAnnoteFile{NoStop}{halliwell09}%
\bibitem{CS10a}%
  \BibitemOpen
  \bibfield{author}{%
  \bibinfo {author} {\bibfnamefont{D.~A.}\ \bibnamefont{Craig}}\ and\ \bibinfo
  {author} {\bibfnamefont{P.}~\bibnamefont{Singh}},\ }%
  \enquote{\bibinfo {title} {Consistent probabilities in wheeler-dewitt quantum
  cosmology},}\  (\bibinfo {year} {2010}),\ \bibinfo {note} {in preparation}%
  \bibAnnoteFile{NoStop}{CS10a}%
\bibitem{GZ05}%
  \BibitemOpen
  \bibfield{author}{%
  \bibinfo {author} {\bibfnamefont{G.}~\bibnamefont{Greenstein}}\ and\ \bibinfo
  {author} {\bibfnamefont{A.~G.}\ \bibnamefont{Zajonc}},\ }%
  \emph{\bibinfo {title} {The quantum challenge: modern research on the
  foundations of quantum mechanics}},\ \bibinfo {edition} {2nd}\ ed.\ (\bibinfo
  {publisher} {Jones and Bartlett, Sudbury},\ \bibinfo {year} {2005})%
  \bibAnnoteFile{NoStop}{GZ05}%
\bibitem{Omnes94}%
  \BibitemOpen
  \bibfield{author}{%
  \bibinfo {author} {\bibfnamefont{R.}~\bibnamefont{Omnes}},\ }%
  \emph{\bibinfo {title} {The interpretation of quantum mechanics}}\ (\bibinfo
  {publisher} {Princeton University Press, Princeton},\ \bibinfo {year}
  {1994})%
  \bibAnnoteFile{NoStop}{Omnes94}%
\bibitem{Griffiths08}%
  \BibitemOpen
  \bibfield{author}{%
  \bibinfo {author} {\bibfnamefont{R.~B.}\ \bibnamefont{Griffiths}},\ }%
  \emph{\bibinfo {title} {Consistent quantum theory}}\ (\bibinfo {publisher}
  {Cambridge University Press, Cambridge},\ \bibinfo {year} {2008})%
  \bibAnnoteFile{NoStop}{Griffiths08}%
\bibitem{CS10b}%
  \BibitemOpen
  \bibfield{author}{%
  \bibinfo {author} {\bibfnamefont{D.~A.}\ \bibnamefont{Craig}}\ and\ \bibinfo
  {author} {\bibfnamefont{P.}~\bibnamefont{Singh}},\ }%
  \enquote{\bibinfo {title} {Consistent probabilities in loop quantum
  cosmology},}\  (\bibinfo {year} {2010}),\ \bibinfo {note} {in preparation}%
  \bibAnnoteFile{NoStop}{CS10b}%
\bibitem{acs:slqc}%
  \BibitemOpen
  \bibfield{author}{%
  \bibinfo {author} {\bibfnamefont{A.}~\bibnamefont{Ashtekar}}, \bibinfo
  {author} {\bibfnamefont{A.}~\bibnamefont{Corichi}},\ and\ \bibinfo {author}
  {\bibfnamefont{P.}~\bibnamefont{Singh}},\ }%
  \bibfield{journal}{%
  \Doi{10.1103/PhysRevD.77.024046}{\bibinfo {journal} {Phys. Rev.}}\ }%
  \textbf{\bibinfo {volume} {D77}},\ \bibinfo {pages} {024046} (\bibinfo {year}
  {2008}),\ \Eprint{http://arxiv.org/abs/0710.3565}{arXiv:0710.3565 [gr-qc]}%
  \bibAnnoteFile{NoStop}{acs:slqc}%
\bibitem{aps:improved}%
  \BibitemOpen
  \bibfield{author}{%
  \bibinfo {author} {\bibfnamefont{A.}~\bibnamefont{Ashtekar}}, \bibinfo
  {author} {\bibfnamefont{T.}~\bibnamefont{Pawlowski}},\ and\ \bibinfo {author}
  {\bibfnamefont{P.}~\bibnamefont{Singh}},\ }%
  \bibfield{journal}{%
  \Doi{10.1103/PhysRevD.74.084003}{\bibinfo {journal} {Phys. Rev.}}\ }%
  \textbf{\bibinfo {volume} {D74}},\ \bibinfo {pages} {084003} (\bibinfo {year}
  {2006}),\ \Eprint{http://arxiv.org/abs/gr-qc/0607039}{arXiv:gr-qc/0607039}%
  \bibAnnoteFile{NoStop}{aps:improved}%
\bibitem{marolf:groupavg}%
  \BibitemOpen
  \bibfield{author}{%
  \bibinfo {author} {\bibfnamefont{D.}~\bibnamefont{Marolf}}}%
   (\bibinfo {year} {1995}),\
  \Eprint{http://arxiv.org/abs/gr-qc/9508015}{arXiv:gr-qc/9508015}%
  \bibAnnoteFile{NoStop}{marolf:groupavg}%
\bibitem{almmt:groupavg}%
  \BibitemOpen
  \bibfield{author}{%
  \bibinfo {author} {\bibfnamefont{A.}~\bibnamefont{Ashtekar}}, \bibinfo
  {author} {\bibfnamefont{J.}~\bibnamefont{Lewandowski}}, \bibinfo {author}
  {\bibfnamefont{D.}~\bibnamefont{Marolf}}, \bibinfo {author}
  {\bibfnamefont{J.}~\bibnamefont{Mourao}},\ and\ \bibinfo {author}
  {\bibfnamefont{T.}~\bibnamefont{Thiemann}},\ }%
  \bibfield{journal}{%
  \Doi{10.1063/1.531252}{\bibinfo {journal} {J. Math. Phys.}}\ }%
  \textbf{\bibinfo {volume} {36}},\ \bibinfo {pages} {6456} (\bibinfo {year}
  {1995}),\ \Eprint{http://arxiv.org/abs/gr-qc/9504018}{arXiv:gr-qc/9504018}%
  \bibAnnoteFile{NoStop}{almmt:groupavg}%
\bibitem{aps}%
  \BibitemOpen
  \bibfield{author}{%
  \bibinfo {author} {\bibfnamefont{A.}~\bibnamefont{Ashtekar}}, \bibinfo
  {author} {\bibfnamefont{T.}~\bibnamefont{Pawlowski}},\ and\ \bibinfo {author}
  {\bibfnamefont{P.}~\bibnamefont{Singh}},\ }%
  \bibfield{journal}{%
  \bibinfo {journal} {Phys. Rev.}\ }%
  \textbf{\bibinfo {volume} {D73}},\ \bibinfo {pages} {124038} (\bibinfo {year}
  {2006}),\ \Eprint{http://arxiv.org/abs/gr-qc/0604013}{arXiv:gr-qc/0604013
  [gr-qc]}%
  \bibAnnoteFile{NoStop}{aps}%
\bibitem{Rovelli:obs}%
  \BibitemOpen
  \bibfield{author}{%
  \bibinfo {author} {\bibfnamefont{C.}~\bibnamefont{Rovelli}},\ }%
  \bibfield{journal}{%
  \Doi{10.1103/PhysRevD.65.124013}{\bibinfo {journal} {Phys. Rev.}}\ }%
  \textbf{\bibinfo {volume} {D65}},\ \bibinfo {pages} {124013} (\bibinfo {year}
  {2002}),\ \Eprint{http://arxiv.org/abs/gr-qc/0110035}{arXiv:gr-qc/0110035}%
  \bibAnnoteFile{NoStop}{Rovelli:obs}%
\bibitem{Marolf:obs}%
  \BibitemOpen
  \bibfield{author}{%
  \bibinfo {author} {\bibfnamefont{D.}~\bibnamefont{Marolf}},\ }%
  \bibfield{journal}{%
  \bibinfo {journal} {Class. Quant. Grav.}\ }%
  \textbf{\bibinfo {volume} {12}},\ \bibinfo {pages} {1199} (\bibinfo {year}
  {1995}),\ \Eprint{http://arxiv.org/abs/gr-qc/9404053}{arXiv:gr-qc/9404053}%
  \bibAnnoteFile{NoStop}{Marolf:obs}%
\bibitem{Dittrich:obs}%
  \BibitemOpen
  \bibfield{author}{%
  \bibinfo {author} {\bibfnamefont{B.}~\bibnamefont{Dittrich}},\ }%
  \bibfield{journal}{%
  \Doi{10.1088/0264-9381/23/22/006}{\bibinfo {journal} {Class. Quant. Grav.}}\
  }%
  \textbf{\bibinfo {volume} {23}},\ \bibinfo {pages} {6155} (\bibinfo {year}
  {2006}),\ \Eprint{http://arxiv.org/abs/gr-qc/0507106}{arXiv:gr-qc/0507106}%
  \bibAnnoteFile{NoStop}{Dittrich:obs}%
\end{thebibliography}%

\end{document}